\documentclass[]{aa}
\usepackage{epsfig}
\begin{document}
\thesaurus{12(...)}

\title{On ``box'' models of shock acceleration and electron 
synchrotron spectra}

\titlerunning{Box models of shock acceleration}

\author{Luke O'C Drury\inst{1}
\and Peter Duffy\inst{2}
\and David Eichler\inst{3}
\and Apostolos Mastichiadis\inst{4}}

\offprints{Luke O'C Drury}

\institute{
School of Cosmic Physics,
Dublin Institute for Advanced Studies,
5 Merrion Square,
Dublin 2,
Ireland
\and 
Department of Mathematical Physics,
University College Dublin,
Dublin 4,
Ireland
\and
Department of Physics,
Ben-Gurion University of the Negev,
Beer-Sheva,
Israel
\and
University of Athens,
Athens,
Greece}

\date{Received 30 January 1999, Accepted 6 May 1999}

\maketitle

%
%
\def\be{\begin{equation}}
\def\ee{\end{equation}}
\def\eqb{\begin{eqnarray}}
\def\eqe{\end{eqnarray}}
\def\racc{r_{\rm acc}}
\def\resc{r_{\rm esc}}
\begin{abstract}
The recent detection of high energy $\gamma$-rays coming
from supernova remnants and active galactic nuclei has revived
interest in the diffusive shock acceleration of electrons.  In the present
paper we examine the basis of the so-called ``box'' model for particle
acceleration and present a more physical version of it.  Using this we
determine simple criteria for the conditions under which ``pile-ups''
can occur in shock accelerated electron spectra subject to synchrotron
or inverse Compton losses (the latter in the Thompson limit). An
extension to include nonlinear effects is proposed.


\end{abstract}

\section{Introduction}

The EGRET detection aboard the Compton Gamma Ray Observatory
of at least two supernova remnants (Esposito \cite{esposito96}) and more than
fifty active galactic nuclei (Thomson et al. \cite{thompson95}) has given
strong evidence of particle acceleration in these objects.
This evidence is strengthened even more by the detection
of SN 1006 (Tanimori \cite{tanimori98}) and the BL Lac objects Mkn 421 
(Punch et al. \cite{punch92}) and Mkn 501 (Quinn et al. \cite{quinn96})
by ground based Cherenkov detectors at TeV energies.

A particularly attractive mechanism for producing the required
radiating high energy particles is the diffusive shock
acceleration scheme, which has already
been put forward to predict TeV radiation from supernova remnants
(Drury, Aharonian \& V\"olk \cite{drury94}, Mastichiadis \cite{masti96}) or 
explain the observed flaring behaviour  
in X-rays and TeV $\gamma$-rays from active galactic nuclei 
(Kirk, Rieger \& Mastichiadis \cite{kirk98}). This scheme was 
originally proposed as the 
mechanism responsible for producing the nuclear cosmic ray component
in shock waves associated with supernova remnants
(Krymsky \cite{krymsky77}, Axford \cite{axford81}). 
Based on this picture many authors 
(Bogdan \& V\"olk \cite{bogdan83}, Moraal \& Axford \cite{moraal83},
Lagage \& Cesarsky \cite{lagage83}, Schlickeiser \cite{schlickeiser84}, 
V\"olk \& Biermann \cite{volk88}, Ball \& Kirk \cite{ball92}, 
Protheroe \& Stanev \cite{protheroe98}) have used, under various guises, 
a simplified but physically intuitive treatment of shock
acceleration, sometimes referred to as a ``box'' model. 

In this paper we examine the underlying assumptions of the ``box'' model 
(\S 2) and we present an alternative more physical
version of it (\S 3). We then include synchrotron and inverse
Compton losses as a means of spectal modification and we 
determine the conditions
under which ``pile-ups'' can occur in shock accelerated spectra
(\S 4). The ``box'' model can also be extended to include the nonlinear 
effect of the particle pressure on the background flow (\S 5).

\section{The ``box'' model of diffusive shock acceleration}

The main features of the ``box'' model, as presented in
the literature (see references above) and exemplified by
Protheroe and Stanev (\cite{protheroe98})) can be summarised as follows.  
The particles being accelerated (and thus
``inside the box'') have differential energy spectrum $N(E)$ and are
gaining energy at rate $\racc E$ but simultaneously escape from the
acceleration box at rate $\resc$. Conservation of particles then
requires
\be
{\partial N\over\partial t}
+{\partial\over\partial E}\left(\racc E N\right) = Q - \resc N
\ee
where $Q(E)$ is a source term combining advection of particles into
the box and direct injection inside the box.

In essence this approach tries to reduce the entire acceleration
physics to a ``black box'' characterised simply by just two rates,
$\resc$ and $\racc$. These rates have of course to be taken from more
detailed theories of shock acceleration (eg Drury \cite{drury91}).  A minor
reformulation of the above equation into characteristic form, \be
{\partial N\over\partial t} + \racc E {\partial N\over \partial E} = Q
- N\left(\resc + \racc + E{\partial\racc\over\partial E}\right) \ee is
useful in revealing the character of the description. This is
equivalent to the ordinary differential equation, \be {d\,N\over d\,t}
= Q - N\left(\resc + \racc + E{\partial\racc\over\partial E}\right)
\ee on the family of characteristic curves described by \be {d\,E\over
d\,t} = \racc E \ee giving the formal solution, 
\eqb 
N(E,t) = \int_0^t
&Q&(t',E')
\nonumber\\
&&exp\left[-\int_{t'}^t \left(\resc + \racc +
E{\partial\racc\over\partial E}\right)\,dt''\right] \,dt'.  
\eqe
The number of particles at energy $E$ and time $t$ in the ``box'' is given
simply by an exponentially weighted integral over the injection rate
at earlier times and lower energies. Of particular interest is the
steady solution at energies above those where injection is occuring
which is easily seen to be a power-law with exponent \be {\partial \ln
N\over\partial \ln E} = - \left( 1 + {\resc\over\racc} +
{\partial\ln\racc\over\partial\ln E}\right).  \ee

At first sight (to one familiar with shock acceleration theory) it
appears odd that the exponent depends not just on the ratio of $\resc$
to $\racc$ but also on the energy dependence of $\racc$. However,
as remarked by Protheroe and Stanev,  the
physically important quantity is not the spectrum of particles inside
the fictitious acceleration ``box'' but the escaping flux of
accelerated particles $\resc N$ and this is a power-law of exponent
\be
{\partial \ln (\resc N)\over\partial \ln E} = 
- \left( 
1 + {\resc\over\racc} + {\partial\ln\racc\over\partial\ln E}
-{\partial\ln\resc\over\partial\ln E} \right)
.
\ee
Thus provided the ratio of $\racc$ to $\resc$ is fixed, 
the power-law exponent of the spectrum of accelerated particles
escaping from the accelerator is determined only by this ratio
whatever the energy dependence of the two rates.

\section{Physical interpretation of the box model}

We prefer a very similar, but more physical, picture of shock
acceleration which has the advantage of being more closely linked to
the conventional theory. For this reason we also choose to work in
terms of particle momentum $p$ and the distribution function $f(p)$
rather than $E$ and $N(E)$. 

The fundamental assumption of diffusive shock acceleration theory is
that the charged particles being accelerated are scattered by magnetic
structures advected by the bulk plasma flow and that, at least to a
first approximation, in a frame moving with these structures the
scattering changes the direction of a particle's motion, but not the
magnitude of its velocity, energy or momentum. If we measure $p$, the
magnitude of the particle's momentum, in this frame, it is not changed
by the scattering and the angular distribution is driven to being very
close to isotropic. However if a particle crosses a shock front, where
the bulk plasma velocity changes abruptly, then the reference frame
used to measure $p$ changes and thus $p$ itself changes slightly. If we have
an almost isotropic distribution $f(p)$ at the shock front where the
frame velocity changes from $\bf U_1$ to $\bf U_2$, then it is easy to
calculate that there is a flux of particles upwards in momentum
associated with the shock crossings of
\eqb
\Phi(p,t) &=& \int p {\bf v\cdot(U_1 - U_2)\over v^2} p^2f(p,t) {\bf v\cdot
n} d\Omega\nonumber\\ 
&=&{4\pi p^3\over 3} f(p,t) {\bf n\cdot(U_1 - U_2)}
\eqe
where $\bf n$ is the unit shock normal and the integration is over all
directions of the velocity vector $\bf v$. Notice that this flux is
localised in space at the shock front and is strictly positive for a
compressive shock structure. 

This spatially localised flux in momentum space is the essential
mechanism of shock acceleration and in our description replaces the
acceleration rate $\racc$. The other key element of course is the loss
of particles from the shock by advection downstream. We note that the
particles interacting with the shock are those located within about
one diffusion length of the shock. Particles penetrate upstream a
distance of order $L_1 = \bf n\cdot K_1\cdot n/n\cdot U_1$ where $\bf K$
is the diffusion tensor and the probability of a downstream particle
returning to the shock decreases exponentially with a scale length of
$L_2 = \bf n\cdot K_2\cdot n/n\cdot U_2$. Thus in our picture we have
an energy dependent acceleration region extending a distance $L_1$
upstream and $L_2$ downstream. The total size of the box is then 
$L(p)\equiv L_1(p)+L_2(p)$. Particles are swept out of this region
by the downstream flow at a bulk velocity $\bf n\cdot U_2$.

Conservation of particles then leads to the following approximate
description of the acceleration,
\be
{\partial\over\partial t}\left[4\pi p^2 fL\right]
+{\partial\Phi\over\partial p}
= Q - {\bf n\cdot U_2} 4\pi p^2 f,
\ee
that is the time rate of change of the number of particles involved 
in the acceleration at momentum $p$ plus the divergence in the accelerated 
momentum flux equals the source minus the flux carried out of the back of 
the region by the downstream flow. The main approximation here is the 
assumption that the same $f(p,t)$ can be used 
in all three terms where it occurs.
In fact in the acceleration flux it is the local distribution at the
shock front, in the total number it is a volume averaged value, and
in the loss term it is the downstream distribution which matters.
Diffusion theory shows that in the steady state all three are equal,
but this need not be the case in more elaborate transport models
(Kirk, Duffy \& Gallant \cite{kirk96}).

Substituting for $\Phi$ and simplifying we get the equation 
\be 
L{\partial f\over\partial t} +
{\bf n\cdot U_1} f(p) + {1\over3} {\bf
n\cdot\left(U_1-U_2\right)} p{\partial f\over\partial p}
= {Q\over 4\pi p^2} 
\ee 
which is our
version of the ``Box'' equation. Note that this, as is readily seen,
gives the well known standard results for the steady-state spectrum
and the acceleration time-scale. In fact our description is mathematically
equivalent to that of Protheroe and Stanev as is easily seen by noting
that
\be
\racc = {{\bf n\cdot(U_1-U_2)}\over 3L},
\quad\resc = {{\bf n\cdot U_2}\over L}, 
\quad N = 4\pi p^2 f L.
\ee
However our version has more physical content, in particular the two
rates are derived and not inserted by hand. It is also important to
note that in our picture the size of the ``box'' depends on the
particle energy.

\section {Inclusion of additional loss processes}

In itself the ``box'' model would be of little interest beyond
providing a simple ``derivation'' of the acceleration time scale. Its
main interest is as a potential tool for investigating the effect of
additional loss processes on shock acceleration spectra. One of the
first such studies was that of Webb, Drury and Biermann (\cite{webb84}) where
the important question of the effect of synchrotron losses was
investigated (see also Bregman et al. \cite{bregman81}).  
An interesting question is whether
or not a ``pile-up'' occurs in the accelerated particle spectrum at
the energy where the synchrotron losses balance the acceleration.
Webb, Drury and Biermann (\cite{webb84}) 
found that pile-ups only occured if the
spectrum in the absence of synchrotron losses (or equivalently at low
energies where the synchrotron losses are insignificant) was harder
than $f\propto p^{-4}$. However Protheroe and Stanev obtain pile-ups
for spectra as soft as $f\propto p^{-4.2}$.

\begin{figure}
\epsfxsize=\hsize
\epsfbox{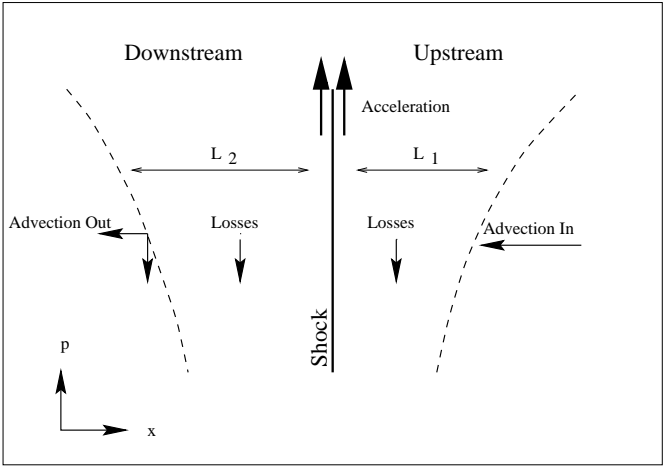}
\caption{Sketch of the acceleration region or ``box'' indicating 
the particle fluxes. The dashed lines indicate the front and back edges of 
the acceleration region.}
\label{boxfigure}
\end{figure}

It is relatively straightforward to include losses of the synchrotron
or inverse Compton type (Thomson regime) in the model.  These generate
a downward flux in momentum space, but one which is distributed
throughout the acceleration region. Combined with the fact that the
size of the ``box'' or region normally increases with energy this also
gives an additional loss process because particles can now fall
through the back of the ``box'' as well as being advected out of it
(see Fig.~1). Note that particles which fall through the front of the
box are advected back into the acceleration region and thus this
process does not work upstream.

If the loss rate is $\dot p = -\alpha p^2$ the basic equation becomes 
\eqb
{\partial\over\partial t}\left[4\pi p^2fL\right]+
{\partial\over\partial p}\left[\Phi - 4\pi p^4 f(p)\alpha L\right]
=\nonumber\\ 
Q - U_2 4\pi p^2 f(p) - 4\pi\alpha p^4 f(p) {dL_2\over dp}
\eqe
This equation is easily generalised to the case of different loss rates 
upstream and downstream. Simplifying equation (12) gives
\eqb
L{\partial f\over\partial t}
+
p{\partial f\over\partial p}
\left[ {U_1-U_2\over 3} -  \alpha pL \right]
+ \nonumber\\
f\left[ U_1 - 4\alpha pL  
- \alpha p^2 {dL_1\over dp} \right]
= {Q\over4\pi p^2}.
\eqe
Note that for convenience we have dropped the explicit vector (and tensor)
notation; all non-scalar quantities are to be interpreted as normal
components, that is $U_2$ means $\bf n\cdot U_2$ etc. Note also that
our model differs from that of Protheroe and Stanev in that
they do not allow for the extra loss process resulting from the energy
dependence of the ``box'' size.

In the steady state and away from the source region this gives
immediately the remarkably simple result for the logarithmic slope of
the spectrum,
\be
{\partial\ln f\over\partial\ln p}
=
-3{\displaystyle{U_1-4\alpha pL
-\alpha p^2 {dL_1\over dp}}\over
U_1 - U_2 - 3\alpha pL}.
\ee
Note that at small values of $p$ we recover the standard result, that the
power-law exponent is $-3 U_1/(U_1 - U_2)$.

\begin{figure}
\epsfxsize=\hsize
\epsfbox{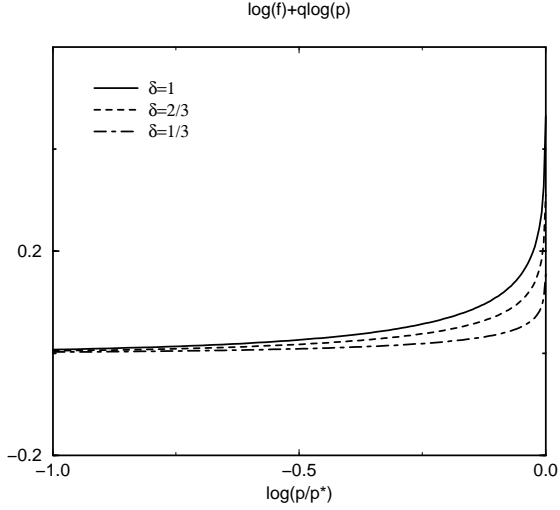}
\caption{Energy spectra for a momentum dependent diffusion coefficient
$\kappa\propto p^\delta$ and a compression ratio of $r=4$ where $q=3r/(r-1)$}
\label{pile1figure}
\end{figure}

\begin{figure}
\epsfxsize=\hsize
\epsfbox{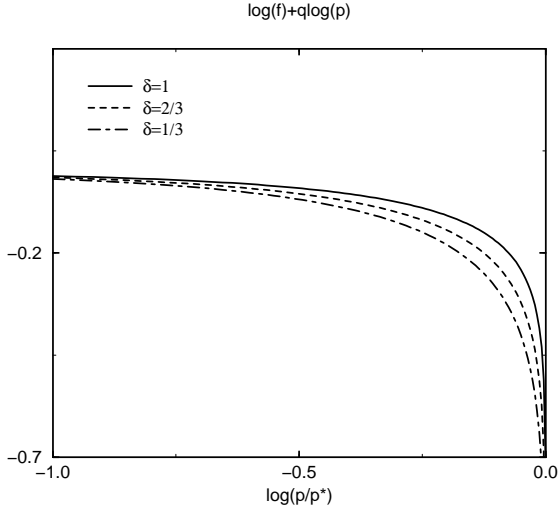}
\caption{As in figure 2 but with $r=3$.}
\label{pile2figure}
\end{figure}

Under normal circumstances both $L_1$ and $L_2$ are monotonically
increasing functions of $p$. Thus both the numerator and denominator
of the above expression, regarded as functions of $p$, have single
zeroes at which they change sign. 
The denominator goes to zero at the critical
momentum
\be
p^* = {U_1 - U_2\over 3\alpha L}
\ee
where the losses exactly balance the acceleration. If the numerator at
this point is negative, the slope goes to $-\infty$ and there is no
pile-up. However the slope goes to $+\infty$ and a pile-up occurs if
\be
U_1 - 4 U_2 + 3\alpha p^2 {dL_1\over dp} > 0
\qquad\hbox{at}\qquad p = p^*.
\ee

In the early analytic work of Webb et al the diffusion coefficient was
taken to be constant, so that $dL_1/dp = 0$
and this condition reduces to $U_1 > 4 U_2$
in agreement with their results. However if, as in the work of
Protheroe and Stanev, the diffusion coefficient is an increasing
function of energy or momentum, the condition becomes less
restrictive. For a power-law dependence of the form $K\propto
p^\delta$
the condition for a pile-up to occur reduces to
\be
U_1 - 4 U_2 + \delta \left(U_1 - U_2\right){L_1\over L_1 + L_2} > 0
\ee
(The equivalent criterion for the model used by Protheroe and Stanev
is slightly different, namely
\be
U_1 - 4 U_2 + \delta \left(U_1 - U_2\right) > 0
\ee
because of their neglect of the additional loss process.)

For the case where $L_1/L_2=U_2/U_1$ and with $\delta=1$ this condition 
predicts that shocks with compression ratios greater than about $r=3.45$ 
will produce pile-ups while weaker shocks will not. In Figures 1 and 2 we 
plot the particle spectra up to $p^*$ for a range of values of $\delta$ 
and with $r=4$ and $r=3$ respectively. 

Thus there is no contradiction between the (exact) results of Webb et
al and those of Protheroe and Stanev; the apparent differences can be
attributed to the energy dependence of the diffusion
coefficient. Indeed, looking at the results presented by Protheroe and
Stanev, it is clear that the pile-ups they obtain are less pronounced
for those cases with a weaker energy dependence.

\section {Nonlinear effects}

At the phenomenological and simplified level of the ``box'' models it
is possible to allow for nonlinear effects by replacing the upstream
velocity with an effective momentum-dependent velocity $U_1(p)$,
reflecting the existence of an extended upstream shock precursor
region sampled on different length scales by particles of different
energies. Higher energy particles, with larger diffusion length
scales, sample more of the shock transition and have larger effective
values of $U_1(p)$; thus $U_1(p)$ must be a monotonically increasing
function of $p$. Repeating the above analysis with a
momentum-dependent $U_1$ the logarithmic slope of the spectrum is in this
case
\be 
{\partial\ln f\over\partial\ln p}
=
-3{\displaystyle{U_1-4\alpha pL+{p\over 3}{dU_1\over dp}
-\alpha p^2 {dL_1\over dp}}\over
U_1 - U_2 - 3\alpha pL}
\ee
with a pile-up criterion of,
\be
U_1(p) - 4U_2 -p{dU_1\over dp} + 3\alpha_1 p^2
{dL_1\over dp} > 0
\qquad\hbox{at}\qquad p = p^*
\ee
We see that whether or not the nonlinear effects assist the formation
of pile-ups depends critically on how fast they make the effective
upstream velocity vary as a function of $p$. By making $U_1(p^*)$
larger they make it easier for pile-ups to occur. On the other hand, if
the variation is more rapid than $U_1\propto p$, the derivative term
dominates and inhibits the formation of pile-ups.

In most cases the shock modification will be produced by the
reaction of accelerated ions, and the electrons can be treated
as test-particles with a prescribed $U_1(p)$. However in a pair
plasma, or if one applies the ``box'' model to the ions themselves,
the effective upstream velocity has to be related to the pressure of
the accelerated particles in a self-consistent way. We require in the 
''box'' model a condition which describes the reaction of the accelerated 
particles on the flow. Throughout the upstream precursor and in the steady 
case both the mass flux, $A\equiv\rho U$, and the momentum flux, $AU+P_C$
are conserved. Here $P_C$ is the pressure contained in energetic particles 
and the gas pressure is assumed to be negligible upstream. At a distance 
$L_1(p)$ upstream only particles with momenta greater than $p$ remain in the 
acceleration region. This suggests that in the ''box'' model the reaction of 
the particles on the flow is described by the momentum flux conservation law
\be
AU_1(p)+\int_p^{p_{\rm max}}4\pi p^2f{pv\over 3}\,dp={\rm constant}
\ee
where $p_{\rm max}$ is the highest momentum particle in the system and
$v$ is the particle velocity corresponding to momentum $p$. 
Differentiating with respect to $p$ gives
\be
A{dU_1(p)\over dp} = 4\pi p^2 f(p) {p v\over 3}.
\ee
With no losses and for $U_1(p)\gg U_2$ we can now recover Malkov's
spectral universality result for strong modified shocks (Malkov,
\cite{malkov98}). In the limit of $U_2= 0$ and $\alpha=0$ the conservation 
equation reduces to the requirement than the upward flux in momentum 
space be constant (equation (9)),
\be
\Phi={4\pi p^3\over 3} f(p) U_1(p)=\Phi_0.
\ee
When combined with equation (22) this gives
\be
U_1 {dU_1\over dp} = {\Phi_0\over A} v = {\Phi_0\over A}
{dT\over dp}
\ee
where we have used the elementary result from relativistic kinematics
that the particle velocity $v$ is the  
derivative of the kinetic energy $T$ with respect to momentum.
Integrating for relativistic particles, $T=pc$,
we get the fundamental self-similar asymptotic solution
found by Malkov,
\be
U_1 = \sqrt{2c\Phi_0\over A} p^{1/2}, \qquad 
f= {3\over4\pi}\sqrt{\Phi_0 A\over 2c}p^{-3.5}.
\ee
If the electrons are test-particles in a shock strongly modified by
proton acceleration, and if the Malkov scaling $U_1\propto p^{1/2}$
holds even approximately, then equation (20) predicts that a  strong 
synchrotron pile-up appears inevitable. 

It is perhaps worth remarking on some peculiarities of Malkov's
solution. Formally it has $U_2 = 0$, all the kinetic energy dissipated
in the ``shock'' is used in generating the upwards flux in momentum
space $\Phi$ and there is no downstream advection. It is not clear
that a stationary solution exists in this case. The problem is that as
$U_2\to 0$ so $L_2\to \infty$ if a diffusion model is used for the
downstream propagation. The solution appears to require some form of
impenetrable reflecting barrier a finite distance downstream if it is
to be realised in finite time. Also, although the accelerated particle
spectrum at the shock is a universal power law, none of these
particles escape from the shock region. From a distance the shock
appears as an almost monoenergetic source at whatever maximum energy
the particles reach before escaping from the system.

The case of a synchrotron limited shock in a pure pair plasma is also
interesting.  Here the upper cut-off is determined not by a free
escape boundary condition but by the synchrotron losses. If most of
the energy dissipated in the shock is radiated this way, the shock
will be very compressive and the downstream velocity $U_2$ negligible
compared to $U_1$. The same caveats about time scales apply as to
Malkov's solution, but again we can, at least as a gedanken
experiment, consider a cold pair plasma hitting an impenetrable and
immovable boundary. In this case, if there is a steady solution, the
upward flux due to the acceleration must exactly balance the
synchrotron losses at all energies. In general it appears impossible
to satisfy both this condition and the momentum balance condition for
$p<p^*$ unless the diffusion coefficient has an artificially strong
momentum dependence. However a solution exists corresponding, in the
box model, to a Dirac distribution at the critical momentum $p^*$.
This steady population of high energy electrons has enough pressure to
decelerate the incoming plasma to zero velocity and radiates away all
the absorbed energy as synchrotron radiation. This extreme form of
pile-up may be of interest as a means of very efficiently converting
the bulk kinetic energy of a cold pair plasma into soft gamma-rays.

\section{Conclusion}

A major defect of all ``box'' models is the basic assumption that all
particles gain and loose energy at exactly the same rate. It is clear
physically that there are very large fluctuations in the amount of
time particles spend in the upstream and downstream regions between
shock crossings, and thus correspondingly large fluctuations in the
amount of energy lost. The effect of these variations will be to smear
out the artificially sharp pile-ups predicted by the simple ``box''
models. However our results are based simply on the scaling with
energy of the various gain and loss processes together with the size
of the acceleration region. Thus they should be relatively robust and
we expect that even if there is no sharp spike, the spectrum will show
local enhancements over what it would have been in the absence of the
synchrotron or IC losses in those cases where our criterion is
satisfied. 

\section{Acknowledgments}

This work was supported by the TMR programme of the EU under contract
FMRX-CT98-0168. Part of the work was carried out while LD was Dozor
Visiting Fellow at the Ben-Gurion University of the Negev; the warm
hospitality of Prof M Mond and the stimulating atmosphere of the
BGU is gratefully acknowledged.


\begin{thebibliography}{}

\bibitem[1981]{axford81}
Axford, W.I. 1981, in Proceedings of 17th Int. Cosmic Ray Conf., 
Paris, 12, 155

\bibitem[1992]{ball92}
Ball, L.T., Kirk, J.G. 1992, ApJ 396, L39


\bibitem[1983]{bogdan83}
Bogdan, T.J., V\"olk, H.J. 1983, A\&A 122, 129

\bibitem[1981]{bregman81}
Bregman, J.N., Lebofsky, M.J., Aller, M.F., Rieke, G.H., Aller, H.D.,
Hodge, P.E., Glassgold, A.E., Huggins, P.J., 1981, Nature 293, 714

\bibitem[1991]{drury91}
Drury, L.O'C. 1991, MNRAS 251, 340

\bibitem[1994]{drury94}
Drury, L.O'C., Aharonian, F., V\"olk, H.J. 1994, A\&A 287, 959

\bibitem[1996]{esposito96}
Esposito, J.A., Hunter, S.D., Kanbach, C., Sreekumar, P. 1996,
ApJ 461, 820

\bibitem[1996]{kirk96}
Kirk, J.G., Duffy, P., Gallant, Y.A. 1996, A\&A 314, 1010

\bibitem[1998]{kirk98}
Kirk, J.G., Rieger, F.M., Mastichiadis, A. 1998, A\&A 333, 452 

\bibitem[1977]{krymsky77}
Krymsky, G.F. 1977, Sov. Phys. Dokl. 22, 327

\bibitem[1983]{lagage83}
Lagage, P.O., Cesarsky, C.J. 1983, A\&A 125, 249

\bibitem[1998]{malkov98}
Malkov, M.A. astro-ph/9807097
 
\bibitem[1996]{masti96}
Mastichiadis, A. 1996, A\&A 305, 53 

\bibitem[1983]{moraal83}
Moraal, H., Axford, W.I. 1983, A\&A 125, 540

\bibitem[1998]{protheroe98}
Protheroe,R.J., Stanev, T.  astro-ph/9808129

\bibitem[1992]{punch92}
Punch, M. et al. 1992, Nature 358, 477

\bibitem[1996]{quinn96}
Quinn, J. et al. 1996, ApJ 456, 83

\bibitem[1984]{schlickeiser84}
Schlickeiser, R. 1984, A\&A 136, 227

\bibitem[1995]{thompson95}
Thompson, D.J. et al. 1995, ApJS 101, 259

\bibitem[1998]{tanimori98}
Tanimori, T. et al. 1998, ApJ 497, L25
 
\bibitem[1984]{webb84}
Webb G.M., Drury L.O'C., Biermann, P.L. 1984, A\&A 137, 185

\bibitem[1988]{volk88}
V\"olk, H.J., Biermann, P.L. 1988, ApJ 333, L65

\end{thebibliography}
\end{document}